\documentclass[11pt,twoside]{article}
\usepackage{asp2010}

\resetcounters

\newcommand{\sla}{\raisebox{-0.10em}{$\stackrel{<}{{\mbox{\tiny $\sim$}}}$}}

\begin{document}

\title{IrOnIc: How to Consider Hundreds of Millions of Iron-Group Lines in NLTE Model-Atmosphere Calculations}

\author{Ellen Ringat}

\affil{Institute for Astronomy and Astrophysics, 
       Kepler Center for Astro and Particle Physics, 
       Eberhard Karls University, 
       Sand 1, 
       D-72076 T\"ubingen, Germany \\ 
       email: gavo@listserv.uni-tuebingen.de}

\begin{abstract}
Iron-group elements have a very high number of atomic levels and an overwhelming number of spectral lines. 
No NLTE model-atmosphere code can cope with these in a classical way. 
A statistical approach was developed over the last decade
to decrease the number of levels and lines to a manageable amount.
The Iron Opacity and Interface (\emph{IrOnIc}) 
calculates sampled cross-sections and model-atom files as input for model-atmosphere computations.
\emph{IrOnIc} is presently transferred into a parallelized code to reduce 
the calculation time to a reasonable value. 
It will be accessible by the public as a service of the German Astrophysical 
Virtual Observatory.
\end{abstract}

\section{Stellar Model-Atmospheres and Statistical Treatment of Iron-Group Lines}
\label{sect:treatment}

Stellar model-atmosphere codes generally solve the radiation-transfer equation together with 
constraint equations like e.g\@. hydrostatic and radiative equilibrium. Each of these 
determines a distinct quantity. E.g\@. the effective temperature $T_\mathrm{eff}$ is obtained by solving 
the radiative equilibrium. For NLTE codes, the Saha-Boltzmann equation is replaced 
by the rate equations which leads to a set of coupled, highly non-linear equations.

A stellar model-atmosphere calculation needs to consider metal-line blanketing to create 
reliable models. Iron-group elements (here Ca-Ni) are very important because they have 
hundreds of thousands of of atomic levels and 
hundreds of millions of respective line transitions. 
Due to their partly filled 3d and 4s atomic sub-shells, 
the levels of iron-group elements are located in very narrow energy range and, thus,
they can be seen as a quasi-continuum. 

With\, classical\, model\, atoms\, \citep{rauchdeetjen_2003}\, for\, the\, iron-group elements, 
the numerical capability even of state-of-the-art NLTE codes like the 
T\"ubingen NLTE\, Model-Atmosphere\, Package\,\, 
\citep[\emph{TMAP}\footnote{http://astro.uni-tuebingen.de/\raisebox{.3em}{\tiny $\sim $}TMAP}\,,][]{werneretal_2003}\, 
is\, exceeded\, by far.\, \emph{TMAP}\,\, has\,\, been\,\, developed\,\, since\,\, the\,\, 1980s\,\, and\,\, is\,\, well\,\, established\,\,  
\citep[e.g\@.\,][]{rauchetal_2007,wassermannetal_2010}. It is suitable for hot, compact objects with 
$20\,000\,\mathrm{K} \,\sla\,T_\mathrm{eff}\,\sla\,200\,000\,\mathrm{K}$, and surface 
gravities $4 \ \sla\  \log g \ \sla\  9$. It can consider opacities from hydrogen 
to nickel \citep{rauch_2003}. For the iron-group elements, a statistical treatment 
enables a consideration of an unlimited number of line transitions. 

\begin{figure}[t]
\setlength{\textwidth}{13.5cm}
\plotone{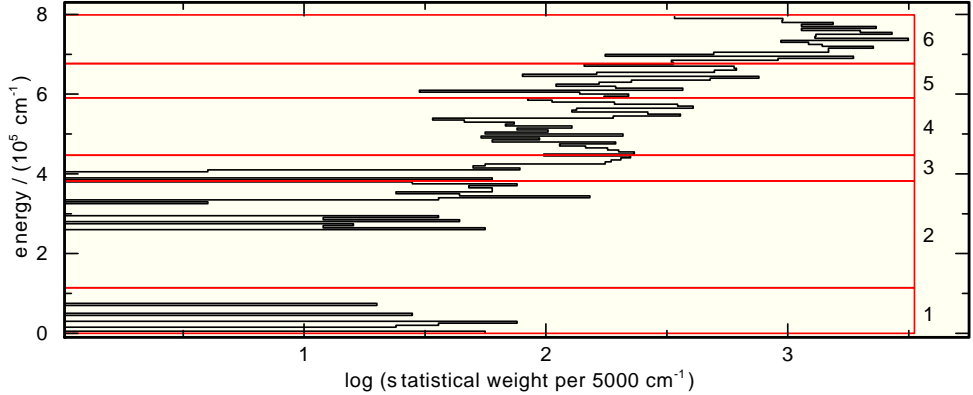}
\caption{Band structure for Fe\,\textsc{vi}. The superlevel number is given on the right.}
\label{fig:band}
\end{figure}

\begin{figure}[t]
\setlength{\textwidth}{13.5cm}
\plotone{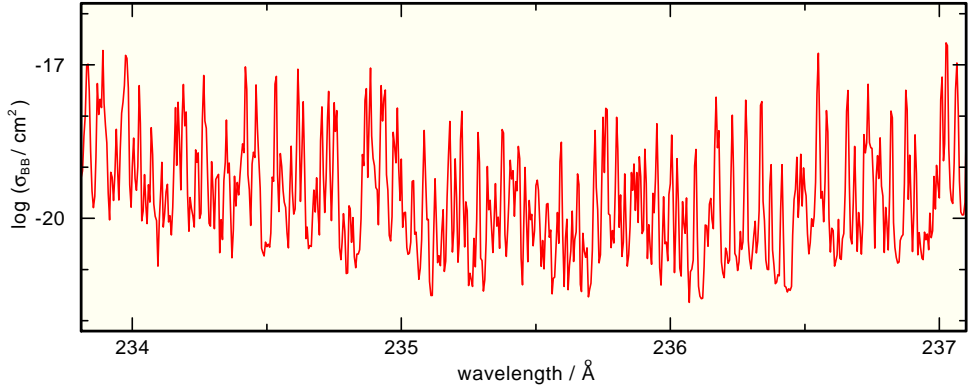}
\caption{Detail of a bound-bound cross-section of Fe\,\textsc{vi} (band 2 to 6, cf\@. Fig.\,\ref{fig:band}).}
\label{fig:cs}
\end{figure}

In order to reduce the quantity of levels and lines for the stellar atmosphere code without 
loosing opacity, \emph{IrOnIc}
has been developed since the late 1990s. The basic idea of this 
code was presented by \citet{anderson_1985, anderson_1989} and adopted for \emph{TMAP} by 
\citet{dreizlerwerner_1992, dreizlerwerner_1993}. It combines all atomic levels of an ion 
into a few so-called superlevels. E.g\@. 24112 levels of Fe\,\textsc{vi} 
are combined into six superlevels (Fig.\,\ref{fig:band}). 
All levels within one superlevel are in LTE relation and 
contribute to its energy and statistical weight. The superlevels are treated by \emph{TMAP} in NLTE. 
Transitions between these superlevels as well as transitions within the bands are allowed. 
They are calculated with an opacity-sampling technique. In this way, every transition 
is represented with its correct strength and position (Fig.\,\ref{fig:cs}).
Source of the atomic data are e.g\@. Kurucz' line lists \citep{kurucz_2009} and the 
Opacity Project \citep{seaton_1994}.

\vspace{-5mm}

\section{The \emph{IrOnIc} Project}
\label{sect:IrOnIc}

Presently, the calculation of cross-sections with \emph{IrOnIc} needs about one to three days. 
As the user cannot start the model-atmosphere calculation until the \emph{IrOnIc} run is performed, 
a new, faster version of this code, using MPI as well as GPU techniques, is developed. 
This happens within the framework of a German 
Astrophysical Virtual Observatory (GAVO\footnote{http://g-vo.org/}) project.

\begin{figure}[t]
\setlength{\textwidth}{12.5cm}
\plotone{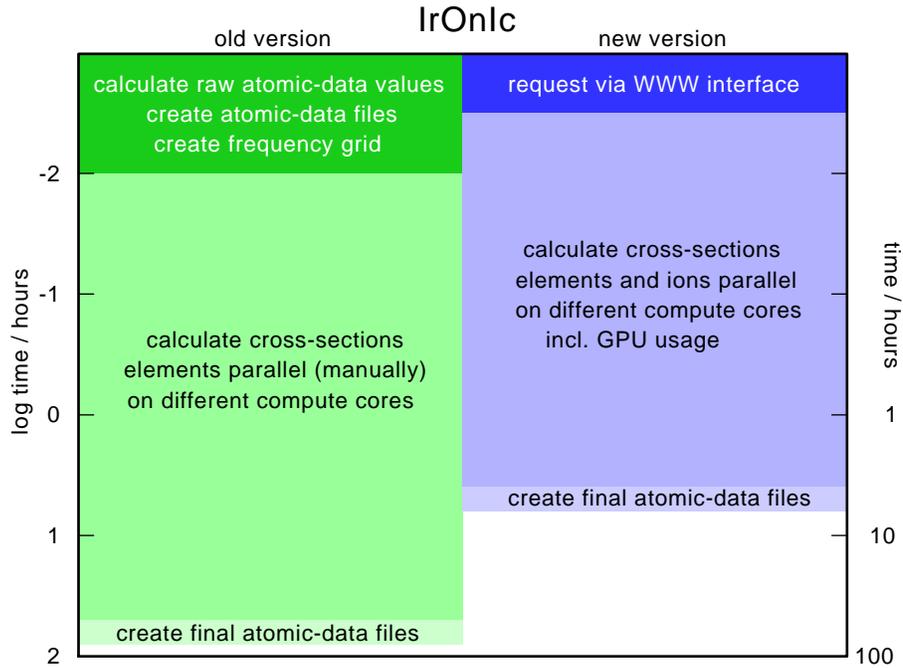}
\caption{Scheme of the old (left) and new (right) \emph{IrOnIc} versions.}
\label{fig:scheme}
\end{figure}

To speed-up the code, three main approaches will be implemented:\\\\
1. With the old \emph{IrOnIc} version, the user has to perform three runs to get the atomic 
data file and cross-sections (Fig.\,\ref{fig:scheme}).  
The new version will be controlled via a web interface. The user only has to enter the 
requested parameters and gets an email notification with a wget command to retrieve cross-sections, 
atomic data, and frequency file. \\
2. The new \emph{IrOnIc} version will use a fixed, fine frequency grid. The cross-sections are 
later interpolated by the model-atmosphere code to the frequency grid of the actual calculation. 
This avoids a complete \emph{IrOnIc} recalculation in case that the frequency grid is changed when 
e.g\@. new atoms are taken into account.\\
3. The code itself will be transferred into a parallel code. The elements as well as their ions 
will run in parallel with MPI parallelization. 
The bottleneck of the calculations are the millions of theoretical line profiles (approximated by 
Voigt profiles). Their calculation presently consumes nearly 90\% of the total calculation time. 
We tested different Voigt function algorithms and got variations up to a factor of ten. Performing 
these calculations on a GPU gives a speed-up of a factor $\approx40$ for the line-profile calculation.\\

\begin{figure}[t]
\setlength{\textwidth}{13.5cm}
\includegraphics[angle=-90,width=13.5cm]{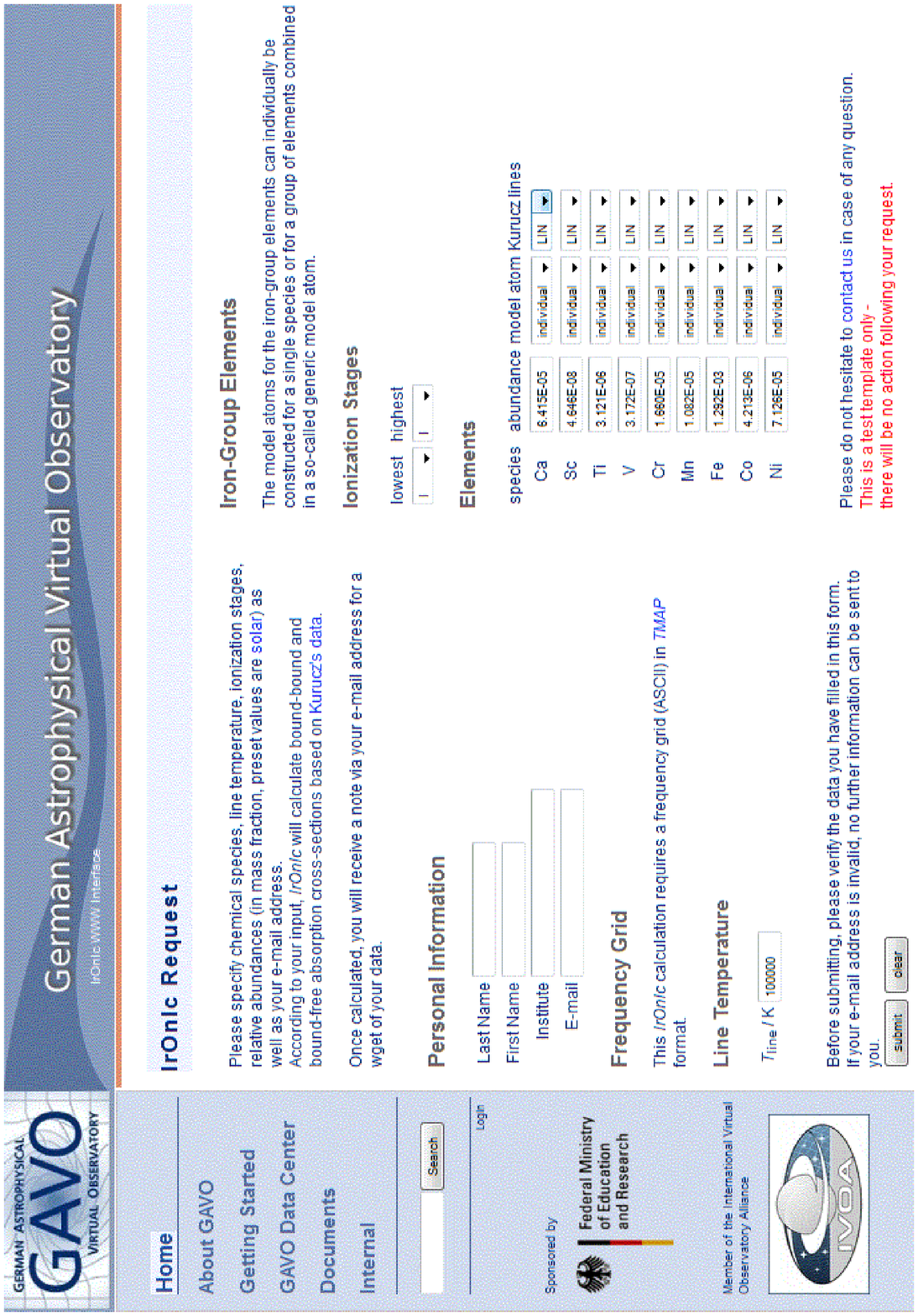}
\caption{The new \emph{IrOnIc} web interface (presently a test version only).}
\label{fig:interface}
\end{figure}

In summary, a new, faster version of the \emph{IrOnIc} code to calculate sampled cross-sections and 
model atoms for iron-group elements is written. As this happens within a \emph{GAVO} project, the 
service will be available to the community. Speed-up will be achieved through parallelization and 
restructuring. The new version is estimated to be a factor of about 20 faster. This  reduces the 
waiting time until the stellar model-atmosphere calculation can start to a value of a few hours instead of days.
In addition, a web interface (Fig.\,\ref{fig:interface}) is created to ensure easy control and 
accessibility, and to provide the data in \emph{VO}-compliant format. 
In this way, any model-atmosphere codes can benefit.

\vfill

\acknowledgements ER is supported by the Federal Ministry of  Education and Research (BMBF) under grant 05A11VTB. 

\bibliography{P129}

\end{document}